\documentclass[twocolumn,nofootinbib,amsmath,amssymb,prd,unsortedaddress]{revtex4-2}

\usepackage{comment}
\usepackage[T1]{fontenc} % if needed
\usepackage{amsmath}
\usepackage{amsthm}
\usepackage{amssymb}
\usepackage{physics}
\usepackage{mathtools}
\usepackage{enumerate}

\usepackage{hyperref}
\usepackage{cleveref}
\usepackage{xcolor}

\usepackage{ulem}

\newcommand{\ud}{{\rm d}}

\newcommand{\phimet}{\Phi}
\newcommand{\psimet}{\Psi}

\newcommand{\To}{\overline T}
\newcommand{\Ti}{\delta T}
\newcommand{\rhoi}{\delta\rho}
\newcommand{\ma}{m_a}
\newcommand{\fa}{f_a}

\newcommand{\mt}{\tilde{m}}
\newcommand{\Tqcd}{T_\textrm{\tiny QCD}}

\newcommand{\thetao}{\overline \theta}
\newcommand{\thetai}{\delta\theta}
\newcommand{\kk}{\boldsymbol{k}}
\newcommand{\x}{\boldsymbol{x}}
\begin{document}

%%%%%%%%%%%%%%%%%%%%%%%%%%%%%%%%%%%%%%%%%%%%%%%%%%%%%%%%%%%%
%           Title, authors, abstract
%%%%%%%%%%%%%%%%%%%%%%%%%%%%%%%%%%%%%%%%%%%%%%%%%%%%%%%%%%%%    	

\title{Axion Perturbations:\\ A General Analytical Treatment}

\author{Itamar J.~Allali}\email{cosmos@itamarallali.com}\affiliation{Department of Physics, Brown University, Providence, RI 02912, USA}
\author{Priyesh Chakraborty}\email{pchakraborty@g.harvard.edu}\affiliation{Department of Physics, Harvard University, Cambridge, MA, 02138, USA}\affiliation{Leung Center for Cosmology and Particle Astrophysics, Taipei 10617, Taiwan}\affiliation{Max Planck--IAS--NTU Center for Particle Physics, Cosmology and Geometry, Taipei 10617, Taiwan}
\author{JiJi Fan}\email{jiji\_fan@brown.edu}\affiliation{Department of Physics, Brown University, Providence, RI 02912, USA}\affiliation{Brown Theoretical Physics Center, Brown University, Providence, RI 02912, USA}
\author{Matthew Reece}\email{mreece@g.harvard.edu}\affiliation{Leinweber Institute for Theoretical Physics at Harvard, Department of Physics, Harvard University, Cambridge, MA, 02138, USA}
 
 \begin{abstract}
Cosmological data provides us two key constraints on dark matter: it must have a particular abundance, and it must have an adiabatic spectrum of density perturbations in the early universe. Many different cosmological scenarios have been proposed that establish the abundance of axion dark matter in qualitatively different ways. In this paper we emphasize that, despite this variety of backgrounds, the perturbations in axion dark matter can be understood from universal principles. How does a feebly interacting axion field acquire perturbations proportional to those of photons? How do the isocurvature power spectrum and non-Gaussianity depend on the background evolution of the universe? We answer these questions for a completely general choice of cosmological background and temperature-dependent axion potential. We show that the most general solution to the axion field equation on super-horizon scales is entirely determined by the family of background solutions for different initial field values $\theta_{\rm ini}$. This holds for both the component in the field perturbation solution contributing to the dark matter isocurvature perturbation (enhanced at late times by the sensitivity of the dark matter abundance to the initial condition, $\partial \Omega_a / \partial \theta_{\rm ini}$, which can be large for initial conditions near the hilltop), and the other component that contributes to the dark matter curvature perturbation. In particular, we explain that an unperturbed axion field in the early universe evolving into one with nontrivial adiabatic perturbations is guaranteed by Weinberg's theorem on adiabatic modes.
These results have been derived before with various assumptions, such as a radiation dominated background or a quadratic potential. Our aim is to give a clear, simple derivation that is manifestly independent of those assumptions, and thus can be applied to any cosmological axion scenario.
 \end{abstract}
\maketitle

%%%%%%%%%%%%%%%%%%%%%%%%%%%%%%%%%%%%%%%%%%%%%%%%%%%%%%%%%%%%%%%%%%%%%%%%

\section{Introduction}

\label{sec:intro}
%%%%%%%%%%%%%%%%%%%%%%%%%%%%%%%%%%%%%%%%%

The QCD axion remains one of the dark matter (DM) candidates with the most significant motivation, and the most motivated such candidate with a non-thermal primary production mechanism. Since the intuitive picture of DM as a thermal relic is quite prevalent in the literature and pedagogical sources, the precise behavior through cosmic time of a non-thermal DM candidate such as the axion can sometimes be elusive. In particular, the process by which the axion excites adiabatic field perturbations in the early universe, arriving at dynamics nearly identical to cold dark matter (CDM), is not as straightforward as the equivalent process for a thermal candidate.\footnote{By CDM we refer to a cosmological fluid with an exact $w = 0$ equation of state, whereas a scalar field like an axion exhibits rapid oscillations in $w$ that average to $w = 0$.} Similarly, the long-term evolution of initial isocurvature fluctuations is not as straightforward as in the CDM case. For these reasons, 
this work expounds upon the generation mechanism and early superhorizon evolution of field fluctuations for the axion and the corresponding density perturbations. The formulae and procedures laid out in this work are generalizable to other scalar DM scenarios beyond the QCD axion, in particular scenarios where the DM does not thermalize with the standard model (SM) bath. 

The QCD axion is a compact pseudo-scalar field $\theta$, postulated to explain the observed non-CP violating nature of QCD. It can arise as a pseudo-Nambu-Goldstone boson for a spontaneously broken, approximate global $U(1)$ Peccei-Quinn (PQ) symmetry~\cite{Peccei:1977hh,Peccei:1977ur,Weinberg:1977ma,Wilczek:1977pj}, or as a mode of an extra-dimensional gauge field~\cite{Witten:1984dg}. The axion DM population is produced via an initial vacuum misalignment in which the scalar is frozen in its potential until the time when the Hubble rate of expansion drops below the effective mass of the field, at which point the field begins to undergo oscillations with an energy density that redshifts like matter \cite{Preskill:1982cy,Abbott:1982af,Dine:1982ah}. This misalignment mechanism proceeds differently depending on whether the axion field is already present as a massless degree of freedom during inflation (pre-inflationary), or emerges from a PQ-breaking phase transition afterwards (post-inflationary). In the pre-inflationary case, the initial phase $\theta_\text{ini}$ is approximately uniform in the observable universe, and thus the axion field is (initially) homogeneous. The post-inflationary case has a complicated cosmology involving production of topological defects, but because the PQ scalar is initially in thermal equilibrium with the SM, the origin of adiabatic fluctuations is less mysterious. For reviews of both cosmological scenarios, see, e.g.,~\cite{Kawasaki:2013ae,Marsh:2015xka,OHare:2024nmr}.

Let us focus on the pre-inflationary axion case. In simple inflationary scenarios, the inflaton exhibits only adiabatic fluctuations, which are then inherited by SM fields following reheating as the universe attains local thermal equilibrium \cite{Hawking:1982cz,Starobinsky:1982ee,Guth:1982ec,Bardeen:1983qw,Gordon:2000hv}. Then, with all fields in equilibrium, the perturbations remain adiabatic and their evolution proceeds according to the usual cosmological perturbation theory and decoupling histories, setting the stage for the cosmic background radiation and evolution of structure that follows \cite{MUKHANOV1992203,Mukhanov:1988jd,Kodama:1984ziu,Ma:1995ey}. However, since the axion is present as a massless field during inflation, entropy-type (isocurvature) inhomogeneities in the axion field are induced, uncorrelated with the adiabatic (curvature) perturbations of the inflaton. For the axion to be a viable DM candidate, these isocurvature perturbations should be small so as not to exceed constraints on DM isocurvature in the cosmic microwave background (CMB) \cite{Planck:2018jri,AtacamaCosmologyTelescope:2025nti, Petretti:2026ayw}. Assuming these inhomogeneities are suppressed, and in the absence of large couplings to the inflaton or the SM particles, the axion field would remain homogeneous in the early universe as it does not come into thermal equilibrium with the rest of the plasma. On the other hand, after the axion begins oscillations, it will behave like DM and thus should have the same inhomogeneities as the adiabatic DM fluctuations (which should be related to the adiabatic perturbations of all other fluids). In fact, as made clear by a theorem of Steven Weinberg regarding the existence of an adiabatic solution to the perturbed Einstein equations at superhorizon scales \cite{Weinberg:2003sw,Weinberg:2003ur,Weinberg:2008zzc}, the initial homogeneous state of the axion is already adiabatic, and it will proceed to remain adiabatic into the regime of its DM-like behavior. Several works in the literature have elaborated on the behavior of the axion and axion-like-particle (ALP) perturbations and the adiabatic nature of the perturbations, with various assumptions about the cosmic background evolution and axion potential~\cite{Axenides:1983hj,Seckel:1985tj,Lyth:1991ub,Marsh:2015xka,Hlozek:2014lca,Hlozek:2017zzf}.

The QCD axion, in addition to being non-thermal DM, has other complications to its early-time dynamics. The axion acquires a potential from QCD dynamics, which gives it an effective mass and higher order couplings. This potential, however, depends on temperature until the completion of the QCD phase transition at a temperature $T_\text{QCD} \sim \mathcal{O}(100 \mbox{ MeV})$. While the evolution of the homogeneous background field remains similar to DM in the presence of this temperature-dependent potential, the perturbations of the field deviate significantly from the DM-like behavior with new contributions arising from temperature fluctuations in the plasma (see, e.g.,~\cite{Ayad:2025awu,Eroncel:2022vjg,Sikivie:2021trt,Kitajima:2021inh} for other sources pointing out this contribution to axion perturbations). This deviation only contributes to the axion field fluctuations in the era prior to the QCD phase transition, but it is not immediately obvious that its effects do not appear at late times by, for instance, spoiling the adiabatic nature of the axion perturbations. 

Both the non-thermal nature of the production of axions and the temperature dependence of the QCD axion potential prior to the QCD phase transition seem at first troublesome for guaranteeing that the axion eventually exhibits an adiabatic fluctuation equivalent to DM. We will see that the adiabaticity of the inhomogeneities survives both of these hurdles, spectacularly guaranteed by Weinberg's theorem. Beyond the adiabaticity of the scalar perturbations, the isocurvature perturbations of the axion also relate to the background evolution. In the context of the QCD axion, the temperature dependent axion potential will give rise to deviations from an otherwise constant entropy perturbation. We will also underscore the influence of the background evolution on the isocurvature spectrum and non-Gaussianity (via the bispectrum).
Though various elements of this discussion are present throughout the literature, a comprehensive explanation of 
the excitation and evolution of the axion's DM-like perturbations in a general cosmological background is lacking. 
In the rest of this work, we will show explicitly that the axion indeed exhibits perturbations which are automatically adiabatic, regardless of its non-thermal production and complicated potential, by first reviewing the theorem by Weinberg followed by an explicit example calculation. We will also comment on the behavior of isocurvature perturbations, which may be excited during inflation, throughout the complex early evolution of the axion field.

%%%%%%%%%%%%%%%%%%%%%%%%
\section{Adiabatic and Entropy Perturbations of Axion}
\label{sec:adiabtic}
%%%%%%%%%%%%%%%%%%%%%%%%

Fields in thermal equilibrium with each other acquire correlated perturbations, so thermally produced dark matter automatically has adiabatic perturbations. The axion is instead produced via the misalignment mechanism and is therefore not immediately required to be adiabatic. Indeed, the standard pre-inflationary axion naturally exhibits fluctuations during inflation which are uncorrelated with the inflaton, resulting in late-time DM isocurvature. In this context, we aim to demonstrate clearly how the axion DM fluid naturally excites an adiabatic mode, despite the complications of the post-inflationary universe such as the QCD phase transition, and that the late-time isocurvature could differ from what is generated by the end of inflation.

The existence and conservation of the adiabatic solution regardless of background cosmological dynamics was shown by Steven Weinberg in a theorem first published in a series of foundational works \cite{Weinberg:2003sw,Weinberg:2003ur} and later expounded upon in his textbook {\it Cosmology} \cite{Weinberg:2008zzc} (for earlier work developing these general solutions, see e.g.~\cite{Mukhanov:1985rz,Hwang:1990am}). Weinberg's theorem guarantees that there always exists a solution to the perturbed scalar field equations in Newtonian gauge (generalizable to other gauges) containing constant scalar curvature $\zeta$ on superhorizon scales with a wavenumber $k$ below the conformal Hubble scale, $k\ll R H$. The perturbed metric in Newtonian gauge is given by
\begin{equation}
     \ud s^2 = -(1+2 \psimet)\ud t^2 + R^2(t)(1-2 \phimet) \ud\boldsymbol{x}^2 \, ,
 \end{equation}
 where $R(t)$ is the scale factor, $H(t)\equiv\dot{R}/R$ is the Hubble rate, and the metric perturbations are observed to be small $\psimet(t, \x), \phimet(t, \x) \ll 1$. 
 The curvature $\zeta$ is
\begin{equation}
    \zeta \equiv -\phimet - H \frac{\rhoi}{\dot {\overline{\rho}}} \, ,
\end{equation}
with $\rhoi$ and $\overline{\rho}$ the total perturbed and background energy densities, and a dot denoting a derivative with respect to $t$.\footnote{The same arguments can also refer to the scalar curvature on comoving spatial surfaces, given by $\mathcal{R}=-\phimet + H \delta u$ with $\delta u$ the total velocity field perturbation, rather than $\zeta$, the scalar curvature on spacelike surfaces of constant energy density. In the superhorizon limit $k \ll R H$, $\zeta=\mathcal{R}$.} Weinberg's argument constructs new inhomogeneous solutions to the equations of motion (up to small-$k$ corrections) by first finding exact new homogeneous solutions and then promoting them to nonzero $k$. 
Given a background solution to the Einstein field equations, the new homogeneous solution is constructed by performing a gauge transformation $x_\mu \mapsto x_\mu + \epsilon_\mu$ which preserves the Newtonian gauge, $\epsilon_\mu=(\epsilon(t), R^2(t)\omega_{ij}x^j)$, on the background solution to yield a scalar fluctuation $\delta s = -\epsilon(t) \dot{\bar{s}}(t)$, where $\delta s$ can be any four-scalar ($\delta\rho$, $\delta p$, $\thetai$, etc.). This can always be done by virtue of diffeomorphism invariance.\footnote{This argument is similar, but not identical, to Goldstone's theorem in QFT, which guarantees the existence of long-wavelength modes when the vacuum state spontaneously breaks a global symmetry. In this context a soft, i.e.~long wavelength, Goldstone mode can locally mimic the shift of a vacuum, which is labeled by a constant phase. In this sense, the Goldstone mode behaves similarly to the cosmological adiabatic mode, and admits a similar construction \cite{Mirbabayi:2016xvc}. In cosmology, the spatially homogeneous background breaks diffeomorphism invariance and likewise engenders the adiabatic mode. In this respect, this mode is sometimes referred to as the Nambu-Goldstone boson of broken \textit{large} diffeomorphisms. A well known consequence is that, similarly to pions, this adiabatic mode also satisfies a host of soft theorems (see e.g.~\cite{Creminelli:2013mca}).}
Demanding that this solution is a physical solution to the field equations fixes $\epsilon_\mu$, and therefore all of the metric and scalar fluctuations.\footnote{This is ensured by demanding the solution holds for nonzero $\kk$, which amounts to satisfying the off-diagonal spatial Einstein equations $k_i k_j(\psimet-\phimet)=-8\pi G k_i k_j \delta \sigma \implies \psimet=\phimet-8\pi G \delta \sigma$ in the presence of anisotropic stress $\delta\sigma$ \cite{Weinberg:2003sw}. In our summary of Weinberg's proof, we take $\delta\sigma=0 \implies \phimet=\psimet$.}
In fact, there are two such solutions, one of which decays in an expanding universe and thus is ignored, while the other is often referred to as the (growing) {\it adiabatic} mode (this solution is also referred to as a curvature perturbation). Upon solving for $\epsilon(t)$ and $\omega_{ij}$, one finds that
\begin{align} \label{eq:epssolution}
    \epsilon(t) &= - \zeta \mu(t), \nonumber \\
    \text{where}\quad\mu(t) &\equiv 
    \frac{1}{R(t)} \int^t R(t')\,\mathrm{d}t'.
\end{align}
Here $\zeta$ is constant, and different choices of the lower limit of integration give rise to solutions differing by a multiple of the decaying mode (and hence are negligibly different at sufficiently late times).
This mode can be promoted to an inhomogeneous mode of long wavelength by extending $\epsilon(t)$ into a function $\epsilon(t,\boldsymbol{x})$. Its Fourier modes, $\epsilon_{\kk}(t)$, are then related to those of the metric perturbation, $\zeta_{\kk}$, by the same proportionality factor $\mu(t)$ as in~\eqref{eq:epssolution}. They solve the equations of motion up to corrections of order $k^2/(R^2 H^2)$.

The (non-decaying) adiabatic solution amounts to the following Fourier modes of scalar metric perturbations
\begin{equation}
\label{eq:weinbergAdiabMetric}          \phimet_{\kk}(t)=\psimet_{\kk}(t)=-\zeta_{\kk} {\dot \mu}(t) = \zeta_{\kk}\left(-1+\frac{H(t)}{R(t)}\int^t R(t')dt'\right).
\end{equation}
The characteristic distinguishing the adiabaticity of this solution is that each species has the same ratio of its energy perturbation to the rate of change of its background energy density $\rhoi_i/\dot{{\overline \rho}}_i$, regardless of whether each species' energy is independently conserved. In fact, the proof of Weinberg's theorem guarantees this relationship for any scalar quantity $s$ \cite{Weinberg:2003sw}, such that
\begin{equation}\label{eq:weinbergAdiabScalar}
    \delta s_{\kk}(t)= -\zeta_{\kk} \mu(t)\dot{\overline{s}}(t) =  -\frac{\zeta_{\kk}\dot{\overline{s}}(t)}{R(t)}\int^t R(t') dt' \, .
\end{equation}
Weinberg's proof guarantees the existence of this adiabatic mode. Data informs us that this mode is a good description of our universe~\cite{Planck:2018jri}. This is an important clue about initial conditions. Single-field inflation always excites the adiabatic mode, but more general physics in the early universe could have led to other solutions. Any deviation from this behavior, then, is non-adiabatic and is labeled an {\it entropy} perturbation, or equivalently an {\it isocurvature} mode. Let us define a generalization of scalar curvature $\zeta_i$ from the perturbed energy of the species $i$ as
\begin{equation}
    \zeta_i \equiv -\phimet - H \frac{\rhoi_i}{\dot{\overline{\rho}}_i} \, .
\end{equation}
Then, the adiabatic mode exhibits equal scalar curvature for each species $\zeta_i = \zeta_j = \ldots$, and any deviation from this behavior contributes to an isocurvature mode. We can define the relative entropy perturbation between two species as
\begin{equation}
    S_{ij} \equiv 3 \left[\zeta_i - \zeta_j\right] \, .
\end{equation}
In the absence of isocurvature, $S_{ij}=0$, the arguments above guarantee that the curvature remains conserved, and thus the isocurvature perturbation remains absent. On the other hand, in the presence of nonzero isocurvature, Weinberg's theorem does not guarantee that the isocurvature necessarily remain constant.

The power and elegance of this theorem is that it applies for arbitrary dynamics of the background cosmology. Examining our case of interest, the QCD axion, we will explicitly show that nontrivial adiabatic perturbations are generated when the axion becomes DM, despite the nontrivial dynamics of the axion. Weinberg's theorem ensures this. At early times, the axion field is frozen, with ${\dot {\overline\theta}}(t) = 0$. The general result~\eqref{eq:weinbergAdiabScalar}, then, tells us that the adiabatic mode is the one with zero perturbations, $\delta \theta_{\kk} = 0$. (This point has been emphasized before, e.g., in section 4.4.1 of the review~\cite{Marsh:2015xka}.) The theorem then guarantees that if the initial condition is in the adiabatic mode, at late times when the axion field oscillates and ${\dot {\overline \theta}}(t) \neq 0$, the solution will still track the adiabatic mode, now with $\delta \theta_{\kk}$ nonzero and proportional to the photon perturbations. Prior to the QCD phase transition, the instanton-induced potential of the axion depends on temperature, modifying the otherwise matter-like behavior of the scalar field and subsequently changing the behavior of the perturbed fields and energies. The remarkable power of Weinberg's theorem is to guarantee that these complications lead to the same late-time solution (on superhorizon scales), provided $\delta \theta_{\kk} = 0$ at sufficiently early times.

In addition, for a pre-inflationary axion some small initial perturbation $\thetai_{\rm ini}$ is expected to be excited during inflation, which will persist as an isocurvature (entropy) perturbation in the post-inflationary universe. In this case, Weinberg's theorem cannot be straightforwardly applied. Especially upon accounting for the QCD phase transition, which generates a coupling between radiation and axions on large-scales, the nonzero entropy mode may evolve nontrivially. We show that while the entropy mode is temporarily excited during the phase transition, the adiabatic mode evolves unscathed, owing primarily to the fact that the axion is an extremely subdominant component of the universe during this period.

In this section, we will present our main derivation in detail. We first derive the equations of motion for both the background and perturbation of the QCD axion DM assuming a general temperature dependent potential and a cosmology dominated by a single fluid with a constant equation of state. In this general setup, we show the exact solution for the axion field perturbation on large scales and identify the adiabatic and isocurvature perturbations. We also comment on non-Gaussianity by computing the bispectrum.

%%%%%%%%%%%%%%%%%%%%%%%%
\subsection{Equations of Motion}
\label{sec:eom}
%%%%%%%%%%%%%%%%%%%%%%%%
The action for the dimensionless axion field $\theta$ with decay constant $f_a$ is
 \begin{equation}
     S = -\int \ud t\,\ud^3 x \, \sqrt{-g}\left[\frac{1}{2} f_a^2 g^{\mu \nu}\partial_\mu \theta \partial_\nu\theta +  V(\theta,T)\right]\,,
 \end{equation} 
 where $g^{\mu\nu}$ is the metric tensor with determinant $g$, and $V(\theta, T)$ is the axion potential. $V(\theta, T)$ could be quite complicated and depends on the temperature, $T$, prior to the QCD phase transition. When $T$ is above the characteristic temperature $T_{\rm QCD}$ of the QCD phase transition, the mass takes an approximately power-law form 
 \begin{equation}\label{eq: axion m(T)}
    \mt(T) \approx \ma \left(\frac{\Tqcd}{T}\right)^n,
\end{equation}
where $m_a$ is the zero-temperature mass and $n \approx 4$. When $T < T_{\rm QCD}$, $\tilde{m}(T) = m_a$. Lattice gauge theory calculations have determined the form of $m(T)$ more precisely, confirming that the exponent $n$ at high temperatures is close to that predicted by the dilute instanton gas approximation~\cite{Borsanyi:2016ksw}. Beyond the quadratic level, the temperature-dependent interaction terms remain to be calculated. These higher-order terms could be important in particular when the initial misalignment angle is large and the axion oscillations explore a wide range of the potential \cite{Turner:1985si, Lyth:1991ub}.  

Let us write the axion field in terms of a (space-independent) background $\overline{\theta}(t)$ and a perturbation $\thetai(t,\x)$,
\begin{equation}
    \theta(t,\x) = \overline{\theta}(t)+\thetai(t,\x) \, .
\end{equation}
We also take the temperature to be perturbed as $T (t, \x) = \overline{T}(t) + \Ti(t, \x)$ with $\overline{T}$ as the background value and $\Ti$ as the perturbation.
 Then, one can expand the action to quadratic order in the complete set of perturbations $\{\thetai,\Ti,\phimet,\psimet\}$ and obtain the equations of motion for $\theta$. Collecting terms by order, we arrive at the zeroth-order evolution equation for the background field $\thetao(t)$ as
 \begin{equation}
    \ddot{\overline{\theta}} + 3 H \dot{\overline{\theta}} + \frac{1}{f_a^2}\frac{\partial \overline{V}}{\partial\overline{\theta}} = 0 \, ,
    \label{eq:background}
\end{equation}
where $\overline{V}=V(\overline{\theta},\overline{T})$. It is well known that the background evolution of the QCD axion is sensitive to the detailed form of $\overline{V}$. The temperature dependence and anharmonic effects could modify the solution of $\overline\theta$ and consequently the predicted relic abundance. 

The first-order perturbation equation for the field perturbation $\thetai(t,\x)$ is
% \begin{equation}
% \delta\ddot{\theta}+3H\delta\dot{\theta}+\left(-\frac{\nabla^2}{R^2} + \frac{1}{f_a^2}\frac{\partial^2 \overline{V}}{\partial{\overline \theta}^2}\right)\thetai
%  -\dot{\overline\theta}(3\dot{\phimet}+\dot{\psimet}) + 2\psimet\frac{1}{f_a^2}\frac{\partial \overline{V}}{\partial\overline \theta}+\frac{1}{f_a^2}\frac{\partial^2 \overline{V}}{\partial \overline \theta\partial \overline T}\Ti =0 \, .
% \end{equation}
\begin{align}
\delta\ddot{\theta}&+3H\delta\dot{\theta}+\left(-\frac{\nabla^2}{R^2} + \frac{1}{f_a^2}\frac{\partial^2 \overline{V}}{\partial{\overline \theta}^2}\right)\thetai
 \\&-\dot{\overline\theta}(3\dot{\phimet}+\dot{\psimet}) + 2\psimet\frac{1}{f_a^2}\frac{\partial \overline{V}}{\partial\overline \theta}+\frac{1}{f_a^2}\frac{\partial^2 \overline{V}}{\partial \overline \theta\partial \overline T}\Ti =0\nonumber \, .
\end{align}
Alternatively, we can write the Fourier transform of the perturbed equation for $\thetai_{\boldsymbol{k}} (t) = \int d^3 \boldsymbol{x}\, \mathrm{e}^{-i{\boldsymbol{k}} \cdot\x} \thetai(t,\x)$. In Fourier space, the perturbation equation becomes
% \begin{equation}
% \delta\ddot{\theta}_{\kk}+3H \delta\dot{\theta}_{\kk}+\left(\frac{k^2}{R^2} + \frac{1}{\fa^2}\frac{\partial^2 \overline{V}}{\partial\overline \theta^2}\right)\thetai_{\kk} \,-\dot{\overline\theta}(3\dot{\phimet}_{\kk}+\dot{\psimet}_{\kk})+2\psimet_{\kk}\frac{1}{\fa^2}\frac{\partial \overline{V}}{\partial\overline \theta}+\frac{1}{\fa^2}\frac{\partial^2 \overline{V}}{\partial \overline \theta\partial \overline T}\delta T_{\kk}  =0 \, ,
% \end{equation}
\begin{align}\label{eq:perteq}
\delta\ddot{\theta}_{\kk}&+3H \delta\dot{\theta}_{\kk}+\left(\frac{k^2}{R^2} + \frac{1}{\fa^2}\frac{\partial^2 \overline{V}}{\partial\overline \theta^2}\right)\thetai_{\kk}\\ &-\dot{\overline\theta}(3\dot{\phimet}_{\kk}+\dot{\psimet}_{\kk})+2\psimet_{\kk}\frac{1}{\fa^2}\frac{\partial \overline{V}}{\partial\overline \theta}+\frac{1}{\fa^2}\frac{\partial^2 \overline{V}}{\partial \overline \theta\partial \overline T}\delta T_{\kk}  =0 \, ,\nonumber
\end{align}
where $k^2\equiv \kk^2$; and $\phimet_{\kk}$,  $\psimet_{\kk}$, and $\delta T_{\kk}$ are the Fourier modes of $\phimet$, $\psimet$ and $\delta T$ with wavenumber $k$. The last three terms in the equation above are all source terms for the axion perturbation. The first two source terms, depending on $\phimet$,  $\psimet$ and their derivatives, originate from the axion field coupling to the metric. They are universally present in perturbation equations of general ALPs, no matter whether the potential is a constant or temperature-dependent. The last term proportional to $\delta T$ takes into account the interaction between the QCD axion and radiation (e.g., gluons). This term is important and has to be included, when the axion potential is temperature dependent, to get the right adiabatic initial conditions, as will be discussed in the next section.\footnote{In Ref.~\cite{Ayad:2025awu}, the source term depending on $\Psi_{\kk}$ is ignored by  an incorrect argument that the background field $\thetao$ makes this term negligible. Then the analysis there only depends on a single source term proportional to $\delta T_{\kk}$. Yet all the source terms could be of the same order at all scales. In particular, one could not get the standard adiabatic initial condition when dropping the source term proportional to $\Psi_k$. } 

The linear-order equations are sufficient to predict the adiabatic and isocurvature power spectrum on large scales. However, we will be interested in determining the axion isocurvature \textit{bispectrum} as well, for which we will need to account for higher order terms in $\thetai_{\kk}$. We will discuss this in detail below.

%%%%%%%%%%%%%%%%%%%%%%%%%%%%%%%%%%%
\subsection{Solution of the Axion Field Perturbation}
%%%%%%%%%%%%%%%%%%%%%%%%%%%%%%%%%%%

We can understand the small-$k$ solution to the axion perturbation equation~\eqref{eq:perteq} quite generally in terms of the family of background solutions ${\bar \theta}(t, \theta_\text{ini}, {\dot \theta}_\text{ini})$, which are parametrized in terms of the initial homogeneous axion value $\theta_\text{ini}$ and its time derivative. Specifically, we make the following assumptions: (i) we take the small-$k$ limit of~\eqref{eq:perteq}, (ii) we take the superhorizon perturbations $\Phi_{\kk}=\Psi_{\kk}$ to be of the adiabatic form in~\eqref{eq:weinbergAdiabMetric}, and (iii) we take the temperature perturbation $\delta T_{\kk}$ (when relevant) to also be of the adiabatic form~\eqref{eq:weinbergAdiabScalar} (we provide justification for these assumptions below). Then, we find the general solution of the form
% \begin{equation}
%  \thetai_{\kk}(t) = A_{\kk} \frac{\partial\overline \theta(t,\theta_\text{ini},{\dot \theta}_\text{ini})}{\partial\theta_\text{ini}} + B_{\kk}  \frac{\partial\overline \theta(t,\theta_\text{ini},{\dot \theta}_\text{ini})}{\partial{\dot \theta}_\text{ini}} - \zeta_{\kk} {\dot {\overline \theta}}(t) \frac{1}{R(t)} \int^tR(t')dt' \, . 
%     \label{eq:generalanalytic}
% \end{equation}
\begin{align}
 \thetai_{\kk}(t) =& A_{\kk} \frac{\partial\overline \theta(t,\theta_\text{ini},{\dot \theta}_\text{ini})}{\partial\theta_\text{ini}} + B_{\kk}  \frac{\partial\overline \theta(t,\theta_\text{ini},{\dot \theta}_\text{ini})}{\partial{\dot \theta}_\text{ini}} \nonumber\\
 &- \zeta_{\kk} {\dot {\overline \theta}}(t) \frac{1}{R(t)} \int^tR(t')dt' \, . 
    \label{eq:generalanalytic}
\end{align}
Here $A_{\kk}$, $B_{\kk}$, and $\zeta_{\kk}$ are constant in time.
This general solution is constructed as follows. The last term, proportional to $\zeta_{\kk}$, follows from Weinberg's theorem and the corresponding general formula~\eqref{eq:weinbergAdiabScalar}, giving the particular solution to~\eqref{eq:perteq}. The remaining terms are obtained by solving the homogeneous operator (i.e., sourceless equation) in~\eqref{eq:perteq}, which can be derived by recognizing the similarities between the homogeneous operators in~\eqref{eq:perteq} and~\eqref{eq:background}, of which $\bar{\theta}(t)$ is a solution. These first two terms, then, give rise to the isocurvature contributions.
Thus, combined with the particular solution (the term proportional to $\zeta_{\kk}$), the general solution in~\eqref{eq:generalanalytic} gives the most general \textit{exact} solution to~\eqref{eq:perteq} in the small-$k$ limit (with the metric and temperature perturbations taken as fixed sources).

Weinberg's formula for the adiabatic part of this general solution, on the other hand, depends crucially on the source terms, and therefore on our assumptions above. In the small-$k$ limit, the source terms depending on the metric perturbations typically take the adiabatic form in~\eqref{eq:weinbergAdiabMetric} when Weinberg's theorem applies. However, in the presence of the evolving axion, the applicability may not be straightforward; for example, in a universe with a power-law expansion $R(t)\propto t^p$, the metric potential sources are strictly constant, but in the presence of the axion they depend on $\thetai_{\kk}$ through Einstein's equations. As a result, it is not obvious that the use of Weinberg's solution~\eqref{eq:generalanalytic} remains valid when the isocurvature contributions are nonzero. Similarly, when the background is dominated by a radiation fluid
which drives the temperature fluctuation $\delta T$
and interacts
with the axion out of equilibrium, 
one must justify that the adiabatic solution for $\delta T_{\kk}$ in equilibrium is valid. We argue in Appendix~\ref{app:constantpotential} that treating both metric perturbations and the radiation perturbation as pure adiabatic perturbations taking the Weinberg solution is, in fact, a good approximation provided that the axion is a subdominant contribution to the total energy density of the universe, as it is around the time of the QCD phase transition when the nonzero adiabatic contributions to $\thetai_{\kk}(t)$ are established. We provide some comments in Appendix~\ref{app:constantpotential} to justify these approximations when the dominant fluid is not radiation, e.g.~if the axion oscillations begin during a period of early matter domination. As a result, we do not need to explicitly solve the equations of motion for $\Phi_{\kk}$, $\Psi_{\kk}$,  and $\delta T_{\kk}$, though they are in principle coupled to~\eqref{eq:perteq}.

In the remainder of this subsection, we will first elaborate on the isocurvature solution and its extension to higher order in perturbations (which is necessary to compute non-Gaussianities in the isocurvature modes). For typical cosmological assumptions, i.e.~a Bunch-Davies initial state for the axion, the initial velocity of the fluctuations is negligible. As we will see below, this allows us to neglect the $B_{\kk}$ terms. Then, we will give a direct demonstration that Weinberg's adiabatic ansatz solves the equations of motion under simplifying assumptions. For this explicit calculation below, we will assume that the universe is dominated by a single fluid and thus evolves according to a power law growth
\begin{equation} \label{eq:Rpowerlaw}
    R(t)\propto t^p
\end{equation} (where for instance, $p=1/2$ corresponds to radiation domination and $p=2/3$ to matter domination). This assumption goes beyond the usual assumption of radiation domination and simplifies the calculation, but it is otherwise unnecessary; we comment later on generalizing these results. In addition, though similarly unnecessary, we assume the background is free of anisotropic stress such that $\phimet=\psimet$, working with $\psimet$ from now on.

Under these simplifying assumptions, we will examine the following solution for $\thetai_{\kk}$ as a special case of~\eqref{eq:generalanalytic}:
\begin{equation}
 \thetai_{\kk}(t) = A_{\kk} \frac{\partial\overline \theta(t,\theta_\text{ini})}{\partial\theta_\text{ini}}  +  t \dot{\overline\theta}(t) \psimet_{\kk} \, . 
 \label{deltathetasoln}
\end{equation}
We will discuss these two terms separately below.

\subsubsection{Isocurvature Solution}

First, we will obtain the general \textit{homogeneous} solution (in the differential equation sense) of the axion fluctuations. We will make minimal assumptions about the background solution, which allow our final result to be applicable beyond the standard misalignment mechanism and include, e.g.~kinetic misalignment \cite{Co:2019jts, Eroncel:2022vjg}.

In general, the axion fluctuations evolve non-linearly due to the non-linear potential, which in turn will produce non-Gaussian statistics for the axion density fluctuations, and therefore also for the isocurvature mode. Note that these non-Gaussianities would arise even if the axion obeyed purely Gaussian statistics by the end of inflation and are therefore distinct from inflationary non-Gaussianities.

In order to obtain the bispectrum, we must first solve the homogeneous equation of motion for the fluctuations (without source terms) beyond linear order. Up to quadratic order we have
\begin{equation}
    \delta\ddot{\theta} + 3H \delta\dot{\theta}+\frac{1}{f_a^2}\bar{V}'' \thetai + \frac{1}{2 f_a^2} \bar{V}''' \thetai^2 =0 \, .
\end{equation}
We can attempt to solve this system perturbatively. A good way to organize the perturbative series is through the initial conditions. To that end, we can write $\thetai = \thetai_{\rm ini} \hat{\thetai}$, with $\thetai_{\rm ini}$ the primordial axion fluctuation generated during inflation.
This parametrization makes it explicit that the non-linear terms are all suppressed due to the small initial fluctuations. Therefore we can write
\begin{equation}
    \hat{\thetai} = \hat{\thetai}^{(1)}+ \thetai_{\rm ini} \hat{\thetai}^{(2)} + \cdots \, ,
\end{equation}
where the initial conditions are enforced at linear order by setting $\hat{\thetai}^{(1)}(t=0)=1$ and $\dot{\hat{\thetai}}^{(1)}(t=0)=0$. The initial conditions for the higher order terms may then be trivial.

At linear order the equation of motion admits a very simple solution in terms of the background solution. It can be verified as $\hat{\thetai}_{\kk}^{(1)}(t)= \frac{\partial \thetao}{\partial \theta_{\rm ini}}$. In fact it is the unique solution which satisfies the initial conditions $\hat{\thetai}_{\kk}(0)=1$ and $\dot{\hat{\thetai}}_{\kk}(0)=0$.\footnote{In general, $\frac{\partial \thetao}{\partial \theta_{\rm ini}}$ and $\frac{\partial \thetao}{\partial \dot{\theta}_{\rm ini}}$ span the space of solutions as linearly independent solutions, as can be verified by computing their Wronskian. As long as $\dot{\hat{\thetai}}_{\kk}(0)=0$, the second solution is not excited. Assuming a Bunch-Davies state for $\thetai$ during inflation, canonical normalization of the massless axion fluctuations ensures $\delta \dot{\theta}\propto a^{-3}$ and is thus negligible by the end of inflation.}

At second order the equations of motion (in position space) are
\begin{equation}
    \ddot{\hat{\thetai}}^{(2)} + 3H \dot{\hat{\thetai}}^{(2)}+\frac{1}{f_a^2}\bar{V}'' \hat{\thetai}^{(2)} + \frac{1}{2 f_a^2} \bar{V}''' (\hat{\thetai}^{(1)})^2 = 0 \, .
\end{equation}
It can be similarly verified by differentiating (\ref{eq:background}) twice that the unique solution to this equation is $\frac{1}{2}\frac{\partial^2\thetao}{\partial \theta_{\rm ini}^2}$. Therefore the complete solution at second order is
\begin{equation}
    \thetai(\x,t) = \thetai_{\rm ini}(\x)\left[\frac{\partial \thetao}{\partial \theta_{\rm ini}} + \frac{1}{2}\thetai_{\rm ini}(\x) \frac{\partial^2 \thetao}{\partial \theta_{\rm ini}^2}\right] \, .
\end{equation}
Indeed, we can recognize this to be a series expansion in the initial data, 
and we can therefore obtain the general solution to $n^{\rm th}$-order 
\begin{equation}
    \thetai(\x,t) = \sum_{n=1}^\infty \left[\frac{1}{n!}\frac{\partial^n \thetao}{\partial \theta_{\rm ini}^n}\right] \thetai_{\rm ini}^n(\x) \, .
\end{equation}
This result can be understood as follows. In the absence of adiabatic perturbations, the initial fluctuations $\thetai_{\rm ini}(\x)$ give a different misalignment angle at each position, so that the non-linear fluctuations are $\thetai(\x,t)=\thetao[\theta_{\rm ini}+\thetai_{\rm ini}(\x),t]-\thetao[\theta_{\rm ini},t]$ which, upon a Taylor expansion, is precisely the expression we have above (such an argument has been used to linear order in e.g.~\cite{Bodas:2025wef}). This works only because we are working in the long-wavelength limit so that spatial derivatives do not affect the time evolution of $\thetai(\x,t)$.

Therefore, the complete non-linear solution for the homogeneous, or isocurvature, field perturbations at all orders of the initial fluctuations $\thetai_{\rm ini}(\x)$ are entirely determined by the background solution. At leading order, we recover the first term in eq.~\eqref{deltathetasoln} by taking the Fourier transform and identifying $A_{\kk} = \thetai_{\rm ini}(\kk)$. 

\subsubsection{Adiabatic Solution}
In the following, we construct the adiabatic solution via an explicit demonstration of Weinberg's theorem.
We now specialize to the power-law scale factor $R(t) \propto t^p$. The second term in eq.~\eqref{deltathetasoln} is, in fact, exactly Weinberg's adiabatic solution.
From \cref{eq:weinbergAdiabMetric}, we obtain
\begin{equation}
    \psimet_{\kk}(t) = \Phi_{\kk}(t) = -\frac{\zeta_{\kk}}{1+p} \,
\end{equation}
and in eq. (\ref{eq:weinbergAdiabScalar})
\begin{equation}
    \thetai_{\kk}(t) = -\frac{\zeta_{\kk}}{1+p}t \,\dot{\overline\theta}(t) = t \dot{\overline \theta}(t)\psimet_{\kk}.
\end{equation}
(Apart from Weinberg's general result, such solutions have appeared in past literature, e.g.,~\cite{Langlois:2004nn}, in specific contexts like a quadratic and temperature-independent potential.) 
Let us verify that the second term, $\thetai_{\kk}= t\dot{\overline\theta}\psimet_{\kk}$, is indeed a (particular) solution to the full perturbation equation. For superhorizon modes, we neglect the $k^2$ term and
take the metric potentials to be constant $\dot{\phimet}_{\kk}=\dot{\psimet}_{\kk}=0$ (though once more, the latter assumption may be relaxed). Plugging this solution in to the left side of eq.~\eqref{eq:perteq}, we have
\begin{align}
(t\dddot{\overline\theta}&+2\ddot{\overline\theta})\psimet_{\kk} + 3 H\psimet_{\kk} (t\ddot{\overline\theta}+\dot{\overline\theta})
    + \frac{1}{\fa^2}\frac{\partial^2 {\overline{V}}}{\partial\thetao^2} (t \dot{\thetao}\psimet_{\kk}) 
    \nonumber\\&+2\psimet_{\kk}\frac{1}{\fa^2}\frac{\partial {\overline{V}}}{\partial\thetao}+\frac{1}{\fa^2}\frac{\partial^2 \overline{V}}{\partial \thetao\partial \To}\delta T_{\kk} = 0\, .
\end{align}
Taking the time derivative of the background equation, eq.~\eqref{eq:background}, we have
\begin{equation}
\dddot{\thetao}+3H\ddot{\thetao}+3\dot{H}\dot{\thetao}+\frac{1}{\fa^2}\left(\frac{\partial^2 \overline{V}}{\partial \thetao^2}\dot{\thetao}+\frac{\partial^2 \overline{V}}{\partial \thetao\partial \To}\dot{\To}\right)=0 \, .
\end{equation}
Combining these two expressions above, we find the left-hand side of Eq.~\eqref{eq:perteq} to be
\begin{align}
 t\psimet_{\kk}&\left(-3\dot{H}\dot{\thetao}-\frac{1}{\fa^2}\frac{\partial^2 \overline{V}}{\partial \thetao\partial \To}\dot{\To}\right)+2 \psimet_{\kk} \ddot{\thetao}+3H \psimet_{\kk} \dot{\thetao} \nonumber\\&+2\psimet_{\kk}\frac{1}{\fa^2}\frac{\partial \overline{V}}{\partial\thetao}+\frac{1}{\fa^2}\frac{\partial^2 \overline{V}}{\partial \thetao\partial \To}\delta T_{\kk} \, = 0 \,.
\end{align}
For $R(t)\propto t^p$, $t\dot{H}=-H$. Thus, using the equation of motion for $\thetao$ in~\eqref{eq:background}, we are left only with
\begin{equation}
- \frac{1}{\fa^2}\frac{\partial^2 \overline{V}}{\partial \thetao\partial \To}\left(t\psimet_{\kk}\dot{\To}-\delta T_{\kk} \right)\, = 0 \,.
\label{eq:laststep}
\end{equation}
However, in the adiabatic mode, the perturbation $\delta T_{\kk}$ is related to $\dot{\To}$ by~\eqref{eq:weinbergAdiabScalar}, which for our background~\eqref{eq:Rpowerlaw} implies that $\delta T_{\kk}/\dot{\To} = t \Psi_{\kk}$. Thus eq.~\eqref{eq:laststep} vanishes and indeed $t\dot{\thetao} \psimet_{\kk}$ is a solution to the full axion field perturbation equation.

Transforming back from the Fourier mode, we could rewrite the adiabatic solution as $\delta \theta (t, \boldsymbol{x})= t \dot{\thetao}(t) \psimet(t, \boldsymbol{x})$. 
We have a few comments on this exact solution at superhorizon scales: 
\begin{itemize}
\item Since the evolution of $\thetao$ depends on $V$, this particular solution tells us that the axion field perturbation $\delta \theta$ also varies for different $V$'s. However, the proportionality relation between $\delta \theta$ and $\dot{\thetao}$ is independent of $V$ and serves as an example of Weinberg's general adiabaticity theorem \cite{Weinberg:2003sw,Weinberg:2003ur,Weinberg:2008zzc}. 
\item The more intuitive picture suggested by the solution is as follows. Initially, the QCD axion is at rest with $\dot{\thetao}=0$, and no field perturbation is generated. Once the axion field starts moving (even before entering the full coherent oscillation phase), the axion field perturbation starts to see the metric and temperature perturbations (correlated as they both arise from the primordial inflaton fluctuations), via the motion of the field. 
\item This solution holds in scenarios beyond the standard misalignment mechanism happening during radiation domination. In particular, it is also a solution in a matter dominated universe. Indeed, one highly motivated scenario to achieve high-scale QCD axion DM with $f_a > 10^{12}$ GeV is to have it start oscillating in an early matter domination phase sourced by a scalar (see, e.g.,~\cite{Steinhardt:1983ia,Lazarides:1987zf,Lazarides:1990xp,Kawasaki:1995vt,Banks:1996ea} for such constructions).\footnote{One small subtlety in this scenario which does not affect our analysis is that in the early matter domination phase, one could ignore the temperature perturbation, while after the scalar reheats, the temperature perturbation is adiabatic.} 
\item While our main physics target is QCD axion DM, the derivation above applies to arbitrary ALP and non-thermal scalar DM, no matter what the potential is. 
\item We assume a single-fluid-dominated universe with a simple power law expansion and constant gravitational perturbations at the superhorizon scale. These are not necessary assumptions. Based on Weinberg's theorem, the adiabatic solution always guarantees the proportionality between $\delta \theta_k$ and $\dot{\thetao} \Psi$. The proportionality factor could be more complicated than $t$, if the background evolution is more complicated. We will discuss this further in Appendix~\ref{app:constantpotential}.
\end{itemize}

%%%%%%%%%%%%%%%%%%%%%%%%%%%%%%%%%%%%
\subsection{Adiabatic Energy Density Perturbation of Axion DM}
%%%%%%%%%%%%%%%%%%%%%%%%%%%%%%%%%%%%

Let us review how the energy and pressure of the adiabatic mode explicitly satisfies the adiabaticity condition. We turn now to the QCD axion as an example with a concrete potential which depends on the field and the temperature. The energy-momentum tensor for the QCD axion is given by
\begin{equation}
     T^\mu_\nu = \fa^2\partial^\mu\theta\partial_\nu\theta - \frac{1}{2}\left[\fa^2\partial^\lambda\theta\partial_\lambda\theta+2V(\theta,T)\right]\delta^\mu_\nu.
\end{equation}
From the $00^{\rm th}$ component, we can expand the energy density of the fluid to the leading two orders and find the background ($\overline{\rho}_\theta$) and first-order perturbation ($\delta \rho_\theta$) of the energy density for axion DM to be 
\begin{align} 
\overline{\rho}_\theta &= \frac{\fa^2}{2} \dot{\thetao}^2 + \overline{V}, \label{eq:rho theta}
\\
\rhoi_\theta &= \fa^2(\dot{\thetao}\delta\dot{\theta}-\dot{\thetao}^2\psimet)+\frac{\partial \overline{V}}{\partial\thetao}\thetai+\frac{\partial \overline{V}}{\partial\To}\Ti.
\end{align}
Then, using the adiabatic solution for the axion field perturbation $\thetai=t\psimet\dot{\thetao}$, we find that the energy density perturbation $\delta \rho_\theta$ is 
\begin{equation}\label{eq:adiabatic delta rho}
   \delta \rho_\theta=-3H\fa^2\dot{\thetao}^2 t \psimet+\frac{\partial \overline{V}}{\partial\To}\delta T\,.
\end{equation}
The time derivative of the axion background energy density is
\begin{equation}
    \dot{\overline{\rho}}_\theta = f_a^2 \dot{\thetao} \,\ddot{\thetao} + \frac{\ud \overline{V}}{\ud t} = f_a^2 \dot{\thetao} \left[-3H \dot{\thetao} - \frac{1}{f_a^2}\frac{\partial \overline{V}}{\partial \thetao}\right] + \dot{\thetao} \frac{\partial \overline{V}}{\partial \thetao} + \dot{\To} \frac{\partial \overline{V}}{\partial \To}\,,
\end{equation}
where using $\dot{\To}=-p  \To/t = - H \To$,
we have
\begin{equation}
    \dot{\overline{\rho}}_\theta = -3Hf_a^2\dot{\thetao}^2- H \To \frac{\partial \overline{V}}{\partial \To} \, .
\end{equation}
Since $H=p/t$ and $\delta T/\To = - p \Psi$, the generalized curvature perturbation for the axion is 
\begin{align}
 \zeta_\theta&=-\phimet-   H\frac{\delta \rho_\theta}{ \dot{\overline{\rho}}_\theta}=-\psimet- H \Psi \frac{-3p f_a^2 \dot{\thetao}^2 -  p \To \frac{\partial \overline{V}}{\partial \To}}{-3Hf_a^2\dot{\thetao}^2- H \To \frac{\partial \overline{V}}{\partial \To}} \nonumber\\&= -(p+1) \Psi
 \, .\end{align}
where in the second equality we have used $\phimet=\psimet$.
On the other hand, for radiation $\delta \rho_\gamma/\overline{\rho}_\gamma = -4 p \Psi$ and $\dot{\overline{\rho}}_\gamma = -4 H \overline{\rho}_\gamma$. Thus 
\begin{equation}
 \zeta_\gamma = - \phimet-   H\frac{\delta \rho_\gamma}{\dot{\overline{\rho}}_\gamma} = -(p+1) \psimet  = \zeta_\theta \, .
\end{equation}
With $\zeta_\gamma=\zeta_\theta$, we have $S_{\theta\gamma}=0$, conclusively showing that $\delta \theta = t \psimet \dot{\thetao}$ is indeed the adiabatic mode. 

Note that for many standard cosmological fluids, $\dot{\overline{\rho}}\propto H \overline{\rho}$, such that the ratio $\rhoi/\dot{\overline{\rho}}$ is proportional to $\rhoi/\overline{\rho}$, and thus often the adiabatic nature of initial conditions for a given fluid is stated as a condition on $\rhoi/\overline{\rho}$. In particular, if a component $\rho_i$ is independently conserved and has a definite equation of state $w_i$, then we have $\dot {\overline{\rho}}_i = -3H(1+w_i) \overline{\rho}_i$. For example, in standard $\Lambda$CDM cosmology, adiabatic fluctuations for CDM are initially given by the density perturbation $\delta_\text{cdm} \equiv \delta\rho_\text{cdm}/\overline{\rho}_\text{cdm}=-3\psimet/2$, while the fluctuation in photon energy density is $\delta_\gamma=-2\psimet$ during radiation domination.  However, when the background energy density evolves nontrivially such that $\dot{\overline{\rho}} \not\propto H \overline{\rho}$, then one must examine the relationship between $\rhoi$ and $\dot{\overline{\rho}}$ to determine the adiabaticity of a solution.

\begin{figure*}
    \centering
    \includegraphics[width=0.8\linewidth]{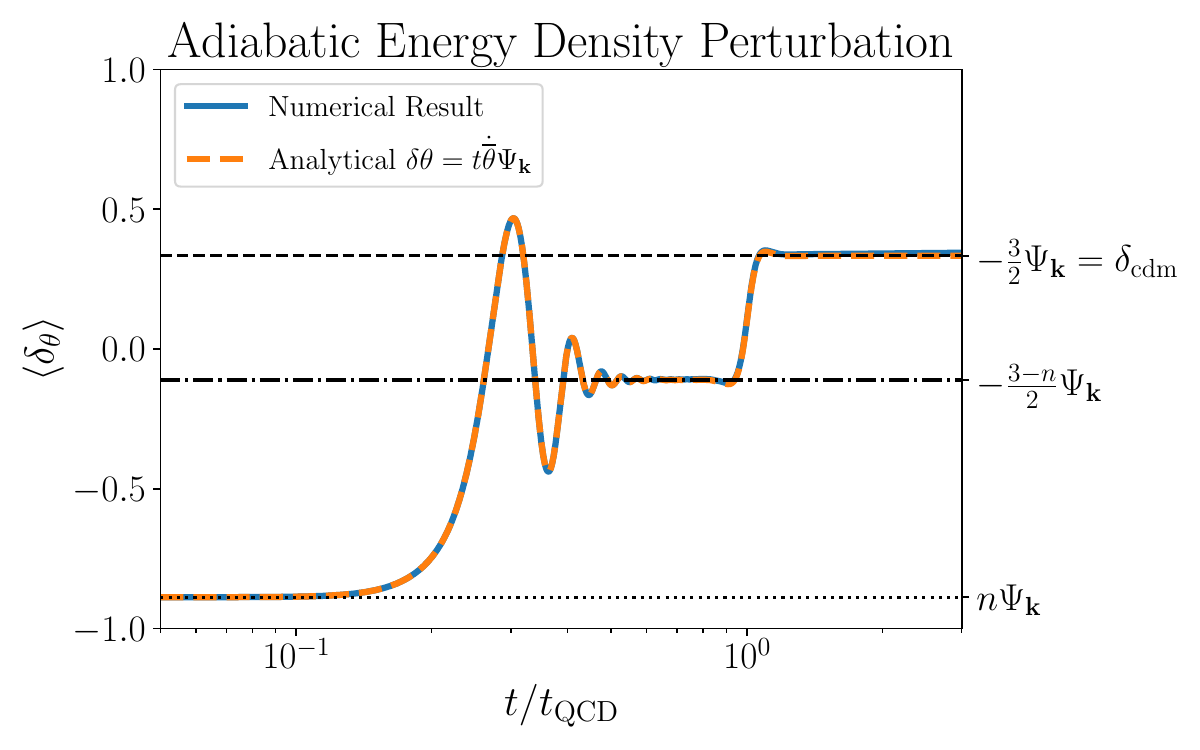}
    \caption{The density perturbation for the QCD axion $\delta_\theta = \delta\rho_\theta/\overline\rho_\theta$, time-averaged by filtering out high-frequency oscillations. A full numerical solution is shown in the blue solid curve, with the analytical solution based on the adiabatic field fluctuation $\thetai = t\dot{\thetao} \psimet_{\kk}$ shown in dashed orange. The black dotted line indicates the initial constant value during the vacuum-dominated phase, while the black dot-dashed line gives the constant about which $\delta_\theta$ oscillates when the QCD axion potential is temperature dependent, and the black dashed line indicates the constant solution after the temperature dependence is gone, identical to the behavior of thermal CDM (when averaging over oscillations). The solutions in this figure assume radiation domination, superhorizon conditions, and the QCD axion potential in \cref{eq: axion m(T)} with $n=4$. 
    }
    \label{fig:adiabatic}
\end{figure*}

In the preceding discussion, we have shown analytically that the adiabaticity of the solution is preserved even when the QCD axion has a temperature-dependent potential. It is useful to examine the behavior of the axion density perturbation $\delta_\theta\equiv  \delta \rho_\theta/\overline{\rho}_\theta$, to see that while it remains adiabatic, its behavior differs from thermal CDM at early times. \Cref{fig:adiabatic} shows both a numerical and analytical solution for $\delta_\theta$. For both cases, high-frequency oscillations are filtered out such that the figure displays oscillation-averaged $\langle\delta_\theta\rangle$. To implement a numerical computation, we assume the QCD axion potential to be quadratic with a temperature-dependent mass given in eq.~\eqref{eq: axion m(T)}, until $t=t_{\rm QCD}$ when the mass becomes constant. Further, we assume radiation domination ($p=1/2$) and superhorizon conditions ($k \ll R H |_{t_{\rm QCD}}$) for the duration of the numerical solution. The analytical solution in the figure corresponds to $\thetai_{\kk}(t) =   t \dot{\overline\theta}(t) \psimet_{\kk} \,$, with $\dot{\overline\theta}(t)$ obtained using the numerical solution for the background $\overline\theta(t)$.

As shown in \cref{fig:adiabatic}, initially the density perturbation is a constant during the vacuum-dominated phase of the axion's evolution. This differs from a vacuum-dominated scalar with constant potential energy density (which would have $\delta_\theta = 0$), as the axion's mass is temperature dependent and thus the potential energy density is time dependent. Combining \cref{eq:adiabatic delta rho}  and \cref{eq:rho theta}, and taking $\dot{\thetao} = 0$, the initial constant is simply $\delta_\theta = \frac{\partial V}{\partial \To} \frac{\delta T}{V} =n\psimet_{\kk}$, where the last equality is valid for the power-law in temperature corresponding to \cref{eq: axion m(T)}, quadratic axion potential, and radiation domination. During this phase, $\delta_\theta$ need not be a constant if the temperature-derivative of the potential is not proportional to the potential itself; indeed for the QCD axion, the potential's temperature dependence could change at early times.  Following the onset of oscillations, $\delta_\theta$ oscillates about a constant which differs from the initial condition for $\delta_\text{cdm}$. Again from \cref{eq:adiabatic delta rho} and \cref{eq:rho theta}, we can see that the initial constant should be $\langle \delta_\theta\rangle=-(3-n)\psimet/2$. Then, after the QCD phase transition is over at $t/t_\text{QCD}=1$ in the figure, the axion potential becomes temperature-independent and it begins to behave as CDM, with the averaged density perturbation $\langle\delta_\theta\rangle=\delta_\text{cdm}=-3\psimet/2$.

\subsection{Isocurvature Energy Density Perturbation of Axion DM}\label{sec: isocurvature}

Let us now return to the homogeneous solution for $\thetai$, which is responsible for the isocurvature component of the DM density fluctuations. Since we will be interested in determining both the isocurvature power spectrum and bispectrum, we will need to determine the axion density fluctuations to second order in the initial conditions. 

One way to obtain the density fluctuations beyond linear order is to use the solution of the axion field perturbations to higher order in the initial data. A shortcut is to recognize that the total energy density is a function of $\theta_{\rm ini}+ \thetai_{\rm ini}$, which means that non-zero initial fluctuations provide a different misalignment angle at each point in space. Therefore one may simply Taylor expand in $\thetai_{\rm ini}(\x)$ to obtain 
\begin{equation}\label{eq:axion_iso_fluct}
    \delta\rho_\theta^{\rm iso}(\x,t) = \frac{\partial \bar{\rho}_\theta}{\partial \theta_{\rm ini}} \thetai_{\rm ini}(\x) + \frac{1}{2} \frac{\partial^2 \bar{\rho}_\theta}{\partial \theta_{\rm ini}^2} \thetai^2_{\rm ini}(\x) + \cdots \, .
\end{equation}
One may verify that the field perturbation 
\begin{equation}
    \thetai(\x,t) = \thetai_{\rm ini}(\x)\left[\frac{\partial \thetao(t)}{\partial \theta_{\rm ini}} + \frac{1}{2} \thetai_{\rm ini}(\x)\frac{\partial^2 \thetao(t)}{\partial \theta_{\rm ini}^2}\right]\,
\end{equation}
reproduces (\ref{eq:axion_iso_fluct}). To check this, we can use the fact that at linear order
\begin{align}
    \delta \rho_\theta^\text{iso,(1)} &= \fa^2\dot{\thetao}\left(A_{\kk} \frac{\partial^2 \thetao}{\partial \theta_{\rm ini} \partial t}\right) + \frac{\partial \overline{V}}{\partial \thetao}A_{\kk} \frac{\partial \thetao}{\partial \theta_{\rm ini}} \nonumber\\&= A_{\kk} \frac{\partial \overline{\rho}_\theta}{\partial \theta_{\rm ini} } = \frac{\partial \overline{\rho}_\theta}{\partial \theta_{\rm ini} }\thetai_\text{ini} \, .
\end{align}
Likewise, one can compute the density fluctuation to quadratic order as well.

Now let us determine the entropy perturbations. At linear order we have
\begin{equation}
    \begin{aligned}
        S_{\theta \gamma}(\kk) &\equiv 3\left[\zeta_{\theta}-\zeta_{\gamma}\right] = -3 H \frac{\delta \rho_{\theta}^{\rm iso}}{\dot{\bar{\rho}}_\theta} \\
        &=-\frac{3H}{\dot{\bar{\rho}}_\theta}\frac{\partial \bar{\rho}_\theta}{\partial \theta_{\rm ini}} \thetai_{\rm ini}(\kk) \, .
    \end{aligned}
\end{equation}
In the absence of temperature dependent effects, after the background oscillations of the axion have begun we have $\dot{\bar{\rho}}_\theta = -3 H \bar{\rho}_\theta$ which yields the simple expression
\begin{equation}\label{eq:energyisopert}
    S_{\theta \gamma} = \frac{\partial \ln \overline{\rho}_\theta}{\partial \theta_{\rm ini} }\thetai_\text{ini} = \frac{\partial \ln \Omega_\theta}{\partial \theta_{\rm ini} }\frac{H_I}{2\pi f_a} \, ,
\end{equation}
where in the last step, we use $\frac{\partial \ln \overline{\rho}_\theta}{\partial \theta_{\rm ini} } =\frac{\partial \ln \Omega_\theta}{\partial \theta_{\rm ini} }$, with $\Omega_\theta$ being the relic abundance of axion dark matter. Strictly speaking, this equality holds only after the axion starts to oscillate in the quadratic part of the potential near the minimum. That is indeed the case for the cosmologically observable modes at large scales. We also use that the primordial axion field fluctuation is set by the inflationary Hubble scale $H_I$: $\delta \theta_{\rm ini} = \frac{H_I}{2\pi f_a}$. To convert this further to CDM isocurvature, one needs to multiply it by the fraction of axion dark matter in the total amount of dark matter, $r S_{\theta \gamma}$ with $r=\Omega_\theta/\Omega_c$. Constraints on dark matter isocurvature from the cosmic microwave background can thus be translated to constraints on the inflationary Hubble scale. \Cref{eq:energyisopert} holds quite generally, and for $\theta_{\rm ini}\sim1$ and $r=1$, this results in a constraint $H_I \lesssim 10^7$ GeV from a combination of \textit{Planck} and ACT-DR6 data (see \cite{Petretti:2026ayw} for detailed updated constraints).

Note that eq.~\eqref{eq:energyisopert} had been derived before~\cite{Kobayashi:2013nva, Lyth:1991ub}, using the $\Delta N$ formalism and during radiation domination, among other assumptions. 
We have shown that this form of the isocurvature mode is more general and follows immediately from the background evolution of the axion. As pointed out in~\cite{Kobayashi:2013nva}, in the large misalignment limit with large $\theta_{\rm ini} \to \pi$, $\delta \rho_\theta^{\rm iso}$ will be enhanced since the axion energy density is more sensitive to fluctuations in the initial misalignment angle in this hilltop region (see \cite{Arvanitaki:2019rax} for a recent thorough study in this limit).

However, accounting for the temperature dependent potential and keeping the axion equation of state arbitrary complicates this expression. Considering the quadratic potential of the QCD axion, we have that $V \propto T^{-2n}$, which yields
\begin{equation}
    \begin{aligned}
        S_{\theta \gamma} &= \left(f_a^2 \dot{\thetao}^2 + \frac{1}{3}\To \frac{\partial \overline{V}}{\partial \To}\right)^{-1}\left(\frac{\partial \bar{\rho}_\theta}{\partial \theta_{\rm ini}} \right)\thetai_{\rm ini} \\
        &= \frac{1}{(1-\tfrac{n}{3})+w_\theta(1+\tfrac{n}{3})}\left(\frac{\partial \log\bar{\rho}_\theta}{\partial \theta_{\rm ini}} \right)\thetai_{\rm ini} \\
        &\approx \frac{2}{1-\tfrac{n}{3}} \frac{\thetai_{\rm ini}}{\theta_{\rm ini}} \, ,
    \end{aligned}
\end{equation}
where in the last line we have used the fact that for a quadratic potential $\bar{\rho}_\theta \propto \theta_{\rm ini}^2$, and taken $w_\theta \approx 0$.\footnote{Note that there is a singularity in this expression for $n=3$. For that potential, the temperature fluctuations compensate for the redshifting of the axion energy density such that $\langle\dot{\rho}_\theta\rangle\approx0$. This can be seen also in \cref{fig:adiabatic}, where $n=3$, drives $\delta_\theta \to 0$. In this case, the curvature perturbation $\zeta_\theta$ is ill-defined, as is the entropy perturbation $S_{\theta\gamma}$.} Choosing, for instance, $n=4$ gives an entropy perturbation $S_{\theta\gamma}=-6\frac{\delta \theta_{\rm ini}} {\theta_{\rm ini}}$, three times larger (and negative) prior to the QCD phase transition when the temperature-dependence is significant, compared to later times when $S_{\theta\gamma}=2\frac{\delta \theta_{\rm ini}} {\theta_{\rm ini}}$.

While this time-dependent shift in the entropy perturbation is of theoretical interest, this behavior is only present at such early times that render it irrelevant for most scales of observational interest (for example the scales that impact the CMB or large-scale-structure observations). Instead, one could readily conceive of scenarios where similar behavior can become observationally relevant. For instance, a dark sector with a phase transition which occurs at a time significantly later than the QCD phase transition might exhibit this shift in entropy perturbations which may impact the CMB power spectra differently than typical constant isocurvature.

\subsection{Non-Gaussianity}

Lastly, we will briefly comment on the isocurvature bispectrum. In general, the axion fluctuations evolve non-linearly due to the non-linear potential, which in turn will produce non-Gaussian statistics for the axion density fluctuations, and therefore also for the isocurvature mode~\cite{Kawasaki:2008sn}. These non-Gaussianities would arise even if the axion obeyed purely Gaussian statistics by the end of inflation and are therefore distinct from inflationary non-Gaussianities. For this calculation, we will turn off the temperature and metric fluctuations as our goal is to calculate the pure isocurvature bispectrum.\footnote{In principle, the entropy mode can be influenced by a coupling between the axion and temperature fluctuations~\cite{Kawasaki:2008sn}.}

To compute the bispectrum we must first determine the comoving curvature of the axion fluid to second order in the density fluctuations which is\footnote{This expression can be found in~\cite{Kobayashi:2013nva}. We have translated their expressions, which are in terms of e-folds, into cosmic time to match our conventions. Recall that $\partial_N = H^{-1}\partial_t$.}
\begin{equation}
    \zeta_\theta^{(2)} = - H \frac{\delta \rho_{\theta}^{(2)}}{\dot{\bar{\rho}}_\theta} +\frac{1}{2}\left(\frac{\dot{H}}{\dot{\bar{\rho}}_{\theta}^2}- H \frac{\ddot{\bar{\rho}}_{\theta}}{\dot{\bar{\rho}}_\theta^3}\right)\left(\delta\rho_\theta^{(1)}\right)^{2} + \frac{H}{\dot{\bar{\rho}}_\theta} \delta\rho_\theta^{(1)}\delta\dot{\rho}_\theta^{(1)} \, ,
\end{equation}
where the superscripts indicate the order at which $\thetai_{\rm ini}$ appears. The second and third terms are more complicated, but can be simplified after some manipulations to yield
\begin{align}
    \frac{1}{2}\left(\frac{\dot{H}}{\dot{\bar{\rho}}_{\theta}^2}- H \frac{\ddot{\bar{\rho}}_{\theta}}{\dot{\bar{\rho}}_\theta^3}\right)&\left(\delta\rho_\theta^{(1)}\right)^{2} + \frac{H}{\dot{\bar{\rho}}_\theta} \delta\rho_\theta^{(1)}\delta\dot{\rho}_\theta^{(1)} \\&= \frac{1}{2 \dot{\bar{\rho}}_\theta}\partial_t\left[\frac{H}{\dot{\bar{\rho}}_\theta}\left(\frac{\partial \bar{\rho}_\theta}{\partial \theta_{\rm ini}}\right)^2\right]\thetai_{\rm ini}^2 \, .\nonumber
\end{align}

In general, as we have previously seen due to temperature dependent effects and the changing axion equation of state, $\dot{\bar{\rho}}_\theta \neq -3 H \bar{\rho}_\theta$, so this expression must be evaluated carefully. In total the curvature perturbation is then
\begin{align}
    \zeta_\theta =& -\frac{H}{\dot{\bar{\rho}}_\theta}\left(\frac{\partial \bar{\rho}_\theta}{\partial \theta_{\rm ini}} \right)\thetai_{\rm ini} \\&- \frac{1}{2}\left[\frac{H}{\dot{\bar{\rho}}_\theta}\frac{\partial^2 \bar{\rho}_\theta}{\partial \theta_{\rm ini}^2} - \frac{1}{\dot{\bar{\rho}}_\theta} \partial_t\left[\frac{H}{\dot{\bar{\rho}}_\theta}\left(\frac{\partial \bar{\rho}_\theta}{\partial \theta_{\rm ini}}\right)^2\right]\right]\thetai_{\rm ini}^2 \, .\nonumber
\end{align}
During QCD phase transition, the bispectrum will also acquire corrections due to the temperature dependent potential, which quickly become irrelevant as the phase transition ends. Therefore, in order to predict the bispectrum relevant for CMB observations, let us assume that the potential is temperature independent and the axion has begun oscillating about the minimum of its potential, and therefore has equation of state $w_\theta=0$.  This expression could then be simplified to 
\begin{equation}
    \frac{1}{2 \dot{\bar{\rho}}_\theta}\partial_t\left[\frac{H}{\dot{\bar{\rho}}_\theta}\left(\frac{\partial \bar{\rho}_\theta}{\partial \theta_{\rm ini}}\right)^2\right] =- \frac{1}{6}\left(\frac{1}{\bar{\rho}_\theta}\frac{\partial \bar{\rho}_\theta}{\partial \theta_{\rm ini}}\right)^2 \, ,
\end{equation}
and the comoving curvature is
\begin{equation}\label{eq:zeta_axion_2nd_gen}
    \zeta_\theta =  \frac{1}{3}\frac{\partial \ln \bar{\rho}_\theta}{\partial \theta_{\rm ini}} \thetai_{\rm ini} + \frac{1}{6}\left[\frac{1}{\bar{\rho}_\theta}\frac{\partial^2 \bar{\rho}_\theta}{\partial \theta_{\rm ini}^2}-\left(\frac{\partial \ln \bar{\rho}_\theta}{\partial \theta_{\rm ini}}\right)^2 \right]\thetai_{\rm ini}^2 \, .
\end{equation}
The DM isocurvature mode $S_{c\gamma}$ up to second order in the axion fluctuations, upon multiplying with the appropriate factor of the relative axion density $r$ is then
\begin{align}
    S_{c \gamma} &= 3 r \zeta_\theta\\ &= r \frac{\partial \ln \bar{\rho}_\theta}{\partial \theta_{\rm ini}} \thetai_{\rm ini} + \frac{r}{2}\left[\frac{1}{\bar{\rho}_\theta}\frac{\partial^2 \bar{\rho}_\theta}{\partial \theta_{\rm ini}^2}-\left(\frac{\partial \ln \bar{\rho}_\theta}{\partial \theta_{\rm ini}}\right)^2 \right]\thetai_{\rm ini}^2 \, , \nonumber
\end{align}
The bispectrum can then be readily calculated to be of the local shape
\begin{equation}
    B_S(k_1, k_2, k_3) = f_{\rm NL}^{(S)}\left[P_S(k_1)P_S(k_2) + {\rm cyc.}\right] \, ,
\end{equation}
where $P_s$'s are power spectra, cyc.~stand for all possible cyclic permutations, and the amplitude is
\begin{equation}
    f_{\rm NL}^{(S)} = -1 + \frac{1}{r \bar{\rho}_\theta}\left(\frac{\partial \ln \bar{\rho}_\theta}{\partial \theta_{\rm ini}}\right)^{-2}\frac{\partial^2 \bar{\rho}_\theta}{\partial \theta_{\rm ini}^2} \, ,
\end{equation}
thus reproducing the expression obtained in \cite{Kobayashi:2013nva}.

%%%%%%%%%%%%%%%%%%%%%%%%%%%%%%%%%%%%%%%
\section{Conclusions}
%%%%%%%%%%%%%%%%%%%%%%%%%%%%%%%%%%%%%%%

In this article, we aim at presenting a clear general derivation for the generation of cosmic perturbations of QCD axion DM in the pre-inflationary scenario. The perturbations consist of two parts, the adiabatic one and the isocurvature one. The most striking feature is that 
irrespective of the complexity of the axion potential and the cosmic background evolution, 
the two types of perturbations in the axion field are entirely determined by the family of axion zero-mode solutions with different initial conditions. 

For the adiabatic perturbation, we apply Weinberg's theorem on the adiabatic mode to show that there is always an exact solution in which the axion field (or energy density) perturbation is proportional to the time derivative of axion background field (or energy density).
For the isocurvature part, we show that the axion field perturbation could be different from the initial perturbation generated during inflation. This results in an enhancement of the late-time energy density fluctuation, as well as non-Gaussianity, by the sensitivity of the DM abundance to the initial conditions,
which can be large when the axion starts oscillating near the hilltop of its potential. This generalizes earlier results~\cite{Lyth:1991ub,Kobayashi:2013nva} to generic background cosmologies (e.g., early matter domination).

While we focus on the QCD axion, our formalism applies to generic ALP or scalar DM, and could aid in understanding the cosmic evolution of non-thermal DM in more complicated scenarios. In addition, some results such as the non-trivial evolution of isocurvature perturbations for a temperature-dependent potential may motivate new cosmological signatures from a dark QCD phase transition.

\section*{Acknowledgements}
We thank Adrienne Erickcek, Paddy Fox, Akshay Ghalsasi, Keisuke Harigaya, Eiichiro Komatsu, Lingfeng Li, Andrew Long, Doddy Marsh, and Sarunas Verner for useful discussions. IJA and JF are supported by the NASA grant 80NSSC22K081 and the DOE grant DE-SC-0010010. PC and MR are supported in part by the DOE Grant DE-SC0013607. This work was performed in part at the Aspen Center for Physics, which is supported by National Science Foundation grant PHY-2210452. JF also acknowledges support of the Institut Henri Poincaré (UAR 839 CNRS-Sorbonne Université), and LabEx CARMIN (ANR-10-LABX-59-01), where part of the work was performed. PC receives support from the Max Planck--IAS--NTU Center, partly funded by NSTC Grant No.~1142923-M-002-011-MY5. Part of this work was conducted using computational resources and services at the Center for Computation and Visualization, Brown University.

\appendix

\section{Validity of a Constant $\Psi$ at Superhorizon Scale}
\label{app:constantpotential}

In the main text, we have identified the adiabatic solution of the axion-radiation system and verified that it holds through non-trivial effects, such as the QCD phase transition, under the mild assumption of a power-law background cosmology. If the number of effective degrees of freedom $g_*$ varies, the power-law background assumption does not hold, and then one needs to rely on the general adiabatic solution in eq.~\eqref{eq:generalanalytic}. Another assumption we have made is that the radiation fluid is unaffected by the axion fluid, and the Newton potential is primarily set by the radiation temperature fluctuations. Here we will verify that this is a good assumption during the QCD phase transition, due to the fact that the axion is an extremely subdominant component of the cosmological fluid and because axion isocurvature fluctuations are observed to be suppressed compared to the adiabatic mode.

In Sec.~\ref{sec:adiabtic}, we work with a power-law expansion history, $R(t) \propto t^p$, and constant gravitational potentials at superhorizon scales. These are valid approximations in a single-fluid dominated universe \cite{Weinberg:2008zzc}. 
Despite corrections from $g_*$, and neglecting the effect of axion isocurvature, we have seen that there is still an exact adiabatic solution for both the gravitational potential and the axion field perturbation, in (\ref{eq:weinbergAdiabMetric}) and (\ref{eq:weinbergAdiabScalar}) respectively.
Recall that to leading order, in a single-fluid dominated universe with $R(t) \propto t^p$, the solution above just reduces to $\Phi_{\kk}=\Psi_{\kk}=- \zeta_{\kk}/(1+p)$, i.e.~a constant and $\delta \theta_{\kk}/\dot{\overline \theta} = t \Psi_{\kk}$, where $\zeta_{\kk}$ is the constant curvature perturbation, as in Sec.~\ref{sec:adiabtic}. 

Now we want to comment on the correction to a constant gravitational potential from the isocurvature perturbation of QCD axion dark matter. Take $\Phi_{\kk}$ as an example. The equation of motion for $\Phi_{\kk}$ at superhorizon scales is 
\begin{equation}
    6 H \dot{\Phi}_{\kk} + 6 H^2 \Phi_{\kk} = -8 \pi G \left[\overline{\rho}_\gamma(t)\delta_{\gamma,\kk} + \overline{\rho}_\theta(t)\delta_{\theta,\kk}\right] \, ,
\end{equation}
where $G$ is the Newton constant, $\delta_\gamma \equiv \delta \rho_\gamma/\overline{\rho}_\gamma$ and $\delta_\theta = \delta \rho_\theta/\overline{\rho}_\theta$,
and $\mathcal{O}(k^2)$ terms are ignored as usual. 
We will write the solution as $\Phi_{\kk} = - \frac{1}{2}\delta_{\gamma,\kk} + \delta\Phi_{\kk}$, where $- \frac{1}{2}\delta_{\gamma,\kk} $ is the leading solution and a constant at superhorizon scale in Newtonian gauge, and $\delta\Phi$ is the correction due to the axion as a subdominant component during radiation domination. At leading order, $3H^2(t) \approx 8\pi G \overline{\rho}_\gamma(t)$. This yields
\begin{align}
    \delta\dot{\Phi}_{\kk} &= -H(t) \delta\Phi_{\kk} -\frac{1}{2}H(t) \frac{\overline{\rho}_{\theta}}{\overline{\rho}_\gamma} \delta_{\theta, \kk}\nonumber \\
    &\Rightarrow \frac{\ud }{\ud t}(R\, \delta \Phi_{\kk}) = -\frac{1}{2}\dot{R}  \frac{\overline{\rho}_\theta}{\overline{\rho}_\gamma} \delta_{\theta,\kk}\, .
\end{align}
The ratio of axion energy density and radiation energy density at an early time, when the scale factor $R \lesssim R_{\rm osc}$,
the scale factor when axion oscillations become cold dark matter,
is
\begin{align}
    \frac{\overline{\rho}_\theta(R)}{\overline{\rho}_\gamma(R)} &= \frac{\overline{\rho}_\theta(R_{\rm osc})}{\overline{\rho}_{\gamma}(R_{\rm osc})} \frac{\overline{\rho}_\gamma (R_{\rm osc})}{\overline{\rho}_\gamma(R)} \frac{\overline{\rho}_\theta(R)}{\overline{\rho}_\theta(R_{\rm osc})}\nonumber\\& \approx \left(\frac{\Omega_\theta}{\Omega_\gamma}\right)R_{\rm osc} \left(\frac{R}{R_{\rm osc}}\right)^{4} \ll 1 \, ,
\end{align}
where $\Omega_{i=\theta, r}$ is the abundance of each species today and we take $\frac{\overline{\rho}_\theta(R)}{\overline{\rho}_\theta(R_{\rm osc})} \sim 1$. For $R > R_{\rm osc}$, the ratio becomes
\begin{equation}
    \frac{\overline{\rho}_\theta(R)}{\overline{\rho}_\gamma(R)} = \left(\frac{\Omega_\theta}{\Omega_\gamma}\right)R \, ,
\end{equation}
which is due to the scaling of energy density for matter and radiation. 
Thus we could estimate that if $R \lesssim R_{\rm osc}$, 
\begin{equation}
    \delta\Phi_{\kk} \lesssim \left(\frac{\Omega_{\theta}}{\Omega_\gamma}\right) \left(\frac{R}{R_{\rm osc}}\right)^3 R \, \delta_{\theta,\kk} \, ,
\end{equation}
and if $R>R_{\rm osc}$ we have
\begin{equation}
    \delta\Phi_{\kk} \sim \left(\frac{\Omega_{\theta}}{\Omega_\gamma}\right) R \, \delta_{\theta, \kk} \, .
\end{equation}
Let us consider the latter case since it sees a larger correction of $\delta \Phi_{\kk}$. The scale factor around the time of the QCD phase transition can be approximated as the ratio of the temperatures at recombination and QCD phase transition, $R = T_{\rm rec}/T_{\rm QCD} \sim (1\,{\rm eV}/100 \,{\rm MeV})\sim 10^{-8}$, and the ratio of abundances today is at most $\Omega_{\theta}/\Omega_\gamma \sim 10^{3}$. 
When $\delta_\theta$ is dominated by the inflationary isocurvature contribution, we have
\begin{align}
    \frac{\delta\Phi}{\Phi} &\sim \left(\frac{\Omega_{\theta}}{\Omega_\gamma}\right) 10^{-8} \frac{\delta\theta_{\rm ini}}{\Phi \theta_{\rm ini}} \sim \left(\frac{\Omega_{c}}{\Omega_\gamma}\right) 10^{-8} \left(\frac{\Omega_{\theta}}{\Omega_c}\right)\frac{\delta\theta_{\rm ini}}{\Phi \theta_{\rm ini}}\nonumber\\&\sim  10^{-5} \sqrt{\frac{A_{\rm iso}}{A_s}} \, ,
\end{align}
where we dropped the subscript $\kk$, $\left(\frac{\Omega_{\theta}}{\Omega_c}\right)\delta\theta_i/\theta_i \sim \sqrt{A_{\rm iso}}$ with $\Omega_c$ the cold dark matter relic abundance today and $A_{\rm iso}$ the amplitude of primordial isocurvature, and $\Phi \sim \sqrt{A_s}$ with $A_s$ the amplitude of the primordial curvature perturbation. Since the isocurvature mode is constrained to be much smaller than the adiabatic mode \cite{Planck:2018jri}, we have roughly $\delta \Phi/\Phi \lesssim 10^{-6}$. This shows that it is a very good approximation to neglect the effects of axion isocurvature onto the adiabatic mode.

Lastly, let us very briefly argue that the backreaction of the axion on the the radiation fluctuations is also suppressed, making the standard adiabatic solution of the temperature fluctuations a valid one to use in the main text. As we have already seen, the axion background energy density and fluctuations both receive important contributions from the temperature dependent potential. One may worry that axion fluctuations may backreact onto radiation as well. In principle, accounting for this energy transfer is complex and depends on particular assumptions about the matter sector. However the total (axion+radiation) stress tensor must be conserved, so we will seek to estimate the explicit temperature dependent contributions to $\rhoi_\theta$, since this can allow us to roughly quantify the effect of the axion on the radiation. Specifically, we must have
\begin{equation}
    \nabla_\mu T^{\mu\nu}_{\theta} = \mathcal{J}^\nu\quad {\rm and} \quad \nabla_\mu T^{\mu\nu}_{\gamma} =- \mathcal{J}^\nu~,
\end{equation}
where the current $\mathcal{J}^\nu$ denotes the flux of energy density between the axion and radiation fluids. We will seek to only estimate the size of $\mathcal{J}^0$. To that end, we can differentiate 
\begin{equation}
    \rhoi_\theta = \fa^2(\dot{\thetao}\delta\dot{\theta}-\dot{\thetao}^2\psimet)+\frac{\partial \overline{V}}{\partial\thetao}\thetai+\frac{\partial \overline{V}}{\partial\To}\Ti~,
\end{equation}
and identify the extra terms generated by the temperature dependent potential. It can be verified that these extra terms are of order $\mathcal{O}(1) \times H \overline{\rho}_\theta\left[ \delta_\theta + \delta_\gamma\right] \sim \mathcal{J}^0$, where the order one factor depends on $n$ via the temperature dependent potential $V\propto T^{-2n}$. Thus the conservation equation for the radiation fluid must be modified such that
\begin{equation}
    \dot{\delta}_\gamma - 4\dot{\Phi} \sim \mathcal{O}(1) \times H \left(\frac{\overline{\rho}_\theta}{\overline{\rho}_\gamma}\right)\left[ \delta_\theta + \delta_\gamma\right]~,
\end{equation}
where we have translated the standard equation for $\rhoi_\gamma$ into one for the radiation overdensity $\delta_\gamma$. As per our estimates with the Newtonian potential, the impact of the axion on the radiation is also extremely suppressed. Well after the QCD phase transition, when $\overline{\rho}_\theta/\overline{\rho}_\gamma$ is not suppressed, $\partial \overline{V}/\partial\To$ is negligible, so there is no interaction at leading order in perturbation theory between the perturbations $\delta \theta$ and $\delta T$. Therefore, the adiabatic solution we rely on is a very good approximation.

If the dominant fluid driving the background evolution is not a radiation-like fluid, many of the arguments above are still valid with some stipulations. The evolution may be driven by some combination of components (a SM bath, a decaying modulus resulting in an early matter-domination, etc.), and in the absence of isocurvature between these components, the adiabatic formulae we have employed still apply. In the presence of some small isocurvature, the above still applies as long as the component with nonzero isocurvature makes up a small enough fraction of the background energy density.

\bibliography{bib}
\end{document}